\documentstyle[multicol,aps,epsf]{revtex}
\begin{document}
%\draft
\title
{Some aspects of functional Bethe ansatz}
\author{You-Quan Li}
\address
{Institut f\"ur Physik, Universit\"at Augsburg, D-86135 Augsburg, Germany\\
and Zhejiang Institute of Modern Physics,
Zhejiang University, Hangzhou 310027, China}
\date{Received 22 Oct 1999}
\maketitle
\begin{abstract}
The validity of Bethe ansatz wave function without the backward scattering
for the problem of impurity in correlated hosts with periodic boundary 
condition
is illustrated by a simple example of quantum mechanics. An long being 
overlooked point in solving Hubbard model by means of Bethe ansatz is 
indicated. A possible repairing on the Hamiltonian is suggested so that
the well known solution is still valid.
\end{abstract}


\begin{multicols}{2}

There has been a long and rich history on the study of integrable models
in condensed matter physics since Bethe solved the one dimensional 
Heisenberg model\cite{Bethe}. One of the important examples is the success
of Bethe ansatz method to Kondo problem. The exact solution of a Kondo
impurity in a free electron host was solved\cite{Andrei,Wiegmann}
by the method in the case of linear dispersion. The Bethe ansatz method 
was recently used to solve a non-magnetic impurity\cite{LiMa}
as well as Kondo impurity\cite{LiBares} in correlated electron hosts 
\cite{Fendley} with quadratic dispersion. In those wave functions, the 
backward scattering waves are not involved\cite{Andrei,Wiegmann,LiMa,LiBares}
because the periodic boundary condition was imposed. It was argued most
recently\cite{Zvyagin} that the reflection matrices for impurities 
at edge\cite{WangVoit} emulate only forward electron-impurity scattering.
In this latter, I show by a simple example of quantum mechanics 
the validity of Bethe ansatz wave function without backward scattering.
By the way, I indicate a long-being overlooked point in solving Hubbard 
model by means of Bethe ansatz, and suggest a choice for repairing the model
Hamiltonian so that the well known solution is still valid.

For simplicity, we place a non-magnetic impurity on a circle. 
The impurity contributes a $\delta$-function type potential 
$V(\theta)=v\delta(\theta)$. For one particle moving on the circle,
if the wave function with backward scattering on the upper patch 
is assumed as
\begin{equation}
\psi_+(\theta>0)=A e^{ik\theta} + B e^{-ik\theta},
\label{eq:WFup}
\end{equation}
then the one on the lower patch should be
\begin{equation}
\psi_-(\theta<0)=A e^{ik2\pi} e^{ik\theta} + B e^{-ik2\pi}e^{-ik\theta}.
\label{eq:WFlow}
\end{equation}   
This is due to the circle geometry (see Fig.1), 
equivalently periodic boundary condition.
At the singularity point $\theta=0$, the wave function is naturally defined
by $(\psi_+(0^+)+\psi_-(0^-))/2$. Now we are in the similar situation 
mathematically for the wave function as for the gauge field in the presence
of magnetic monopoles. As is known that a discontinuity relation at
$\theta=0$ for the derivatives of wave function can be obtained from the 
Sch\"odinger equation by using Gaussian theorem at the neighborhood of 
the $\delta$-function singular point, namely,
\begin{equation}
\psi'_+(0^+) -\psi'_-(0^-)=v[\psi_+(0^+)+\psi_-(0^-)],
\label{eq:discontinuity}
\end{equation}
where the prime stands for derivative. Substituting the wave functions 
(\ref{eq:WFup}) and (\ref{eq:WFlow}) into eq.(\ref{eq:discontinuity}),
one can obtain two possible solutions, one is $B/A=-\exp(ik2\pi)$ and
the other is 
\begin{equation}
e^{ik2\pi}=-\frac{v-ik}{v+ik}
\label{eq:secular}
\end{equation}

Because $v=0$ is the case of a particle moving on a circle without impurity,
the coefficient $A$ and $B$ should obviously be independent. The former
is not a reasonable solution. If we temporally impose the 
continuity condition for the wave function at $\theta=0$, we would get
$B/A=\exp(ik2\pi)$. This contradicts not only with the $v=0$ argument, but
also disagrees with the solution of (\ref{eq:discontinuity}). Differing from 
the relation (\ref{eq:discontinuity}) which can be derived from Schr\"odinger
equation, the continuity condition is not a consequence of the Schr\"odinger
equation. We argue that the  continuity condition should no more be imposed
at present $\delta$-function singularity point otherwise the present quantum
mechanical
problem would have no solution. Therefore eq.(\ref{eq:secular}), which is 
consistent for both $v=\infty$ (i.e., 
$\psi(0):=(\psi_+(0^+)+\psi_-(0^-) )/2=0)$ and $v=0$, 
is the secular equation for the system. 

Now let us observe the above example without the help of backward scattering.
According to the strategy\cite{LiMa,LiBares} we define the wave function
piece-wisely on the ``left'' and ``right'' regions adjacent to the 
impurity,
\begin{equation}
\Psi(x)=\left\{\begin{array}{ll}
               \psi_+(x)=A_+ e^{ikx}  & x>0, \\
               \psi_-(x)=A_- e^{-ikx} & x<0.
               \end{array}  
        \right.
\label{eq:WFnoback}
\end{equation}
The discontinuity relation (\ref{eq:discontinuity}) gives rise to
\begin{equation}
A_-=-\frac{v-ik}{v+ik}A_+.
\label{eq:}
\end{equation}
From the periodic boundary condition 
$\Psi(x+L)=\Psi(x)$, explicitly here $\psi_+(x+L)=\psi_-(x)$,
we obtain once again the same secular equation (\ref{eq:secular}) 
as long as $L=2\pi$, the length of the circle.
For $v\rightarrow\infty$, the wave function defined at the 
point $x=0$ satisfactorily vanishes
which concises  with physical picture. The case of $v=0$ gives 
$A_+=A_-$ which is also of consistency.

If the impurity is a Kondo impurity 
$V(x)=(J S\cdot S_{im}+V)\delta(x)$ the same discussion can be made as
long as we introduce a two component wave function. We will not show 
these for the sake of saving pages.

For a Kondo impurity in $N$ corelated electron host with quadratic dispersion,
the secular equation for the spectrum was solved\cite{LiMa,LiBares} on the 
basis of Bethe ansatz by using the piece-wisely defined wave function 
without the backward scattering waves. 
One can see that the secular equation for $N=1$ 
reduces to the result of the above examples.
Conversely, we can imagine a particle moving on the background of 
the other $N-1$ particles besides the impurity. If employing the picture
of the above example, we know that the particle will undergo $N-1$
phase shifts after moving around the circle once 
due to the scattering with the other
``background'' potentials caused by the $N-1$ particles. Then we
can phenomenally understand the product of the $N-1$ factors
in the Bethe ansatz equation
for boson system. For fermion system, however, it is too complicated to
deduce the Bethe ansatz equation via a vivid description. Whence 
the literature of Bethe ansatz strategy is necessary and useful. 

One can revisit the problem by taking account the backward 
scattering in the Bethe ansatz wave function, i.e.,
\begin{equation}
\psi(x)=\sum_{P\in\cal{W}_B}A(P;Q^{(i)})e^{i(Pk|Qx)} \, 
         {\rm for }\, x\in{\cal C}(Q^{(i)}).
\label{eq:BAWF}
\end{equation}
where $\cal{W_B}$ stands for the Weyl group of $B_N$ Lie algebra;
$a:=(a_{Q1}, a_{Q2}, \cdots, a_{QN})$, $a_j$ 
denotes the spin component of the $j$th particle; 
$Pk$ represents the image of a given $k:=(k_1, k_2, \cdots, k_N )$ by 
a mapping $P \in\cal{W}_B $, i.e., either permutating or adding a minus sign;
$(Pk | Qx) = \sum_{j=1}^{N} (Pk)_j (Qx)_j$ and
${\cal C}(Q^{(i)}):=\{ x | (Qx)_1<(Qx)_2<...<(Qx)_i<0<...<(Qx)_N, Q\in S_N \}$.
Although there are more coefficients $A$'s in this case, 
the Schr\"odinger equation is sufficiently satisfied 
by the same electron-impurity S-matrix. 
%at every patch and each neighborhood of the barriers 
%between different patches. 
Of cause, the argument $(Pk)_i$ in the S-matrix 
$S^{Qi,0}[(Pk)_i]$ takes not only any of the $k$'s in
$\{ k_1, k_2, ...,k_N \}$ but also those with an additional minus sign.
However, due to the property
$S^{Qi,0}[-(Pk)_i]=\left( S^{Qi,0}[(Pk)_i] \right)^{-1}$, one is able to
obtain the same set of independent equations for the spectrum as in 
ref.\cite{LiBares} at the integrable sector. For open boundary
condition, of cause, the backward scattering wave is necessary.
Away from the integrable sector, the backward scattering
plays an essential role. The dynamical backward scattering in a
system of nonintegrable Kondo impurity was shown \cite{DHLee}
to influence the properties of the system. Although the existence of 
impurity breaks infinitesimal translation symmetry, the symmetry of 
translation with a finite distance $L$ remains for periodic 
boundary condition\cite{LiBares}.
This may help to understand the validity of Bethe ansatz wave function
without backward scattering for an integrable Kondo impurity. 

Let us turn to a long-being overlooked point in solving
Hubbard model\cite{LiebWu},
\begin{equation}
H=\sum_{i,a}-t(C^+_{i\,a}C_{i+1\,a}+C^+_{i+1\,a}C_{i\,a})
   +U\sum_{i}n_{i\uparrow} n_{i\downarrow}.
\label{eq:Hubbard}
\end{equation}
The spin states are denoted either by $a=+1/2, -1/2$ or by
$\uparrow$ and $\downarrow$. The correct first quantized version of 
the Hamiltonian (\ref{eq:Hubbard}) in the Hilbert space of 
$N$ particles, in stead,  should be
\begin{equation}
H=-\sum_{j=1}^{N}\Delta_j + U\sum_{i,j}\delta_{-a_i, a_j}\delta_{x_i,x_j}.
\label{eq:NpHamiltonian}
\end{equation}
The second term is spin-dependent which had not been noticed 
for a long period, 
so the result of ref. \cite{Yang} can not be employed to present
case directly. The original model, however, can be repaired.
The antisymmetry property of fermion permutation always requires
the wave function to be null when $x_i=x_j$ and $a_i=a_j$.
This can be explicitly verified even if the wave function is piece-wisely
defined. Consequently, the Hamiltonian (\ref{eq:Hubbard}) 
is allowed to have an additional term
$V\sum_{i,j}\delta_{a_i,a_j}\delta_{x_i, x_j}$ for arbitrary parameter $V$
without changes in physical observables. 
Thus, strictly speaking, the Lieb-Wu solution\cite{LiebWu} was solved from 
the Hamiltonian (\ref{eq:Hubbard}) after repairing the second term to be
\begin{equation}
U\sum_{i,a,b}n_{i\,a} n_{i\,b}.
\end{equation}
Fortunately, the spectrum obtained in ref.\cite{LiebWu} is still valid
without changes due to the above mentioned equivalence relation.    

The work is supported by AvH-Stiftung, 
also supported by NSFC-19975040 and EYFC98.

\end{multicols}
\newpage

\begin{figure}
\setlength{\unitlength}{1mm}
\begin{picture}(100,120)(-66,-16)
\linethickness{1pt}
\put(30.2,30){\circle*{2.2}}
\put(28.9,31.5){\vector(0,1){2.6}}
\put(29,31.5){\vector(0,1){2.95}}
\put(29.1,31.5){\vector(0,1){2.6}}
\put(30.9,34){\vector(0,-1){2.6}}
\put(31,34){\vector(0,-1){2.95}}
\put(31.1,34){\vector(0,-1){2.6}}
\put(28.9,25.8){\vector(0,1){2.6}}
\put(29,25.8){\vector(0,1){2.95}}
\put(29.1,25.8){\vector(0,1){2.6}}
\put(30.9,28.5){\vector(0,-1){2.6}}
\put(31,28.5){\vector(0,-1){2.95}}
\put(31.1,28.5){\vector(0,-1){2.6}}
\linethickness{0.8pt}
\put(0,30){\dashbox{1.0}(30,0)}
\bezier{100}(0,0)(11.78,0)(21.2,8.8)
\bezier{100}(21.2,8.8)(30,18.22)(30,30)
\bezier{100}(30,30)(30,41.78)(21.2,51.2)
\bezier{100}(21.2,51.2)(11.78,60)(0,60)
\bezier{100}(0,60)(-11.78,60)(-21.2,51.2)
\bezier{100}(-21.2,51.2)(-30,41.78)(-30,30)
\bezier{100}(-30,30)(-30,18.22)(-21.2,8.8)
\bezier{100}(-21.2,8.8)(-11.78,0)(0,0)
%\thinlines
\bezier{20}(0,1)(11.38,1)(20.4,9.3)
\bezier{20}(20.4,9.3)(29,18.05)(29,30)
\bezier{20}(29,30)(29,41.3)(20.4,50.7)
\bezier{20}(20.4,50.7)(11.38,59)(0,59)
\bezier{20}(0,59)(-11.38,59)(-20.4,50.7)
\bezier{20}(-20.4,50.7)(-29,41.3)(-29,30)
\bezier{20}(-29,30)(-29,18.05)(-20.4,9.3)
\bezier{20}(-20.4,9.3)(-11.38,1)(0,1)
%\thinlines
\bezier{25}(0,-1)(12.38,-1)(22,8.3)
\bezier{25}(22,8.3)(31,18.39)(31,30)
\bezier{25}(31,30)(31,42.26)(22,51.7)
\bezier{25}(22,51.7)(12.18,61)(0,61)
\bezier{25}(0,61)(-12.18,61)(-22.51,51.7)
\bezier{25}(-22.51,51.7)(-31,42.26)(-31,30)
\bezier{25}(-31,30)(-31,18.39)(-22,8.3)
\bezier{25}(-22,8.3)(-12.38,-1)(0,-1)
\end{picture}
\caption{The circle is separated by an impurity into upper patch $\theta>0$
and lower patch $\theta<0$, on which the wave function of a single particle
is defined piece-wisely. Both forward and backword waves are indicated by
dot-lines. The relations between the coefficients in eq.(\ref{eq:WFup}) and
eq.(\ref{eq:WFlow}) are obvious from the figure}
\label{fig:circle}
\end{figure}
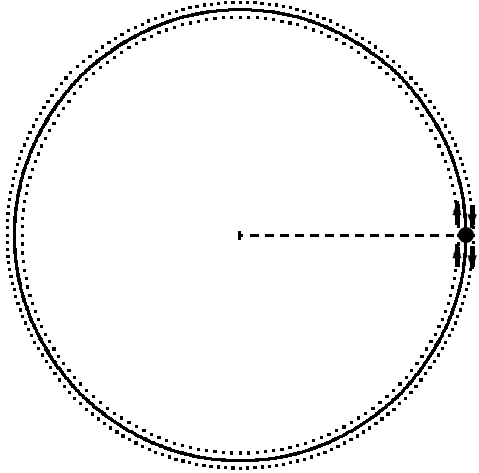
\end{document}